\documentclass[onecolumn,12pt]{article}
\usepackage{graphicx}
\newcommand\simlt{\hspace{0.3em}\raisebox{0.4ex}{$<$}\hspace{-0.75em}\raisebox{-.7ex}{$\sim$}\hspace{0.3em}}

\newcommand{\beginsupplement}{%
        \setcounter{table}{0}
        \renewcommand{\thetable}{S\arabic{table}}%
        \setcounter{figure}{0}
        \renewcommand{\thefigure}{S\arabic{figure}}%
     }

\begin{document}
\baselineskip 14pt

\title{Distortion of Magnetic Fields in a Starless Core: \\
Near-Infrared Polarimetry of FeSt 1-457} 
\author{Ryo Kandori$^{1}$, Motohide Tamura$^{1,2,3}$, Nobuhiko Kusakabe$^{1}$, Yasushi Nakajima$^{4}$\\
Jungmi Kwon$^{5}$, Takahiro Nagayama$^{6}$, Tetsuya Nagata$^{7}$, Kohji Tomisaka$^{2}$, \\
and \\
Ken'ichi Tatematsu$^{2}$\\
{\small 1. Department of Astronomy, The University of Tokyo, 7-3-1, Hongo, Bunkyo-ku, Tokyo 113-0033, Japan}\\
{\small 2. National Astronomical Observatory of Japan, 2-21-1 Osawa, Mitaka, Tokyo 181-8588, Japan}\\
{\small 3. Astrobiology Center of NINS, 2-21-1, Osawa, Mitaka, Tokyo 181-8588, Japan}\\
{\small 4. Hitotsubashi University, 2-1 Naka, Kunitachi, Tokyo 186-8601, Japan}\\
{\small 5. Institute of Space and Astronautical Science, Japan Aerospace Exploration Agency,}\\
{\small 3-1-1 Yoshinodai, Chuo-ku, Sagamihara, Kanagawa 252-5210, Japan}\\
{\small 6. Kagoshima University, 1-21-35 Korimoto, Kagoshima 890-0065, Japan}\\
{\small 7. Kyoto University, Kitashirakawa-Oiwake-cho, Sakyo-ku, Kyoto 606-8502, Japan}\\
{\small e-mail: r.kandori@nao.ac.jp}}
\maketitle
\abstract{\bf Magnetic fields are believed to play an important role in controlling the stability and contraction of dense condensations of gas and dust leading to the formation of stars and planetary systems. 
In the present study, the magnetic field of FeSt 1-457, a cold starless molecular cloud core, was mapped on the basis of the polarized near-infrared light from 185 background stars after being dichroically absorbed by dust aligned with the magnetic field in the core. A distinct ``hourglass-shaped" magnetic field was identified in the region of the core, which was interpreted as the first evidence of a magnetic field structure distorted by mass condensation in a starless core. The steep curvature of the magnetic field lines obtained in the present study indicates that the distortion was mainly created during the formation phase of the dense core. The derived mass-to-magnetic flux ratio indicates that the core is in a magnetically supercritical state. However, the stability of the core can be considered to be in a nearly critical state if the additional contributions from the thermal and turbulent support are included. Further diffusion of the magnetic field and/or turbulent dissipation would cause the onset of dynamical collapse of the core. The geometrical relationship between the direction of the magnetic field lines and the elongation of the core was found to be in good agreement with the theoretical predictions for the formation of Sun-like stars under the influence of a magnetic field. 
}


\vspace*{0.3 cm}

\section{Introduction}  
The characteristics of newborn stars are thought to be determined by the physical properties of the nursing molecular cloud core prior to the onset of gravitational collapse, and the magnetic field pervading the core is believed to play an important role in controlling the stability and contraction of the core (e.g., Shu, Adams, \& Lizano 1987). 
It is therefore important to investigate the magnetic field structure affecting the cores of molecular clouds in the starless phase in order to clarify the initial conditions of star formation. 
However, such investigations are hampered by the difficulty of observing magnetic phenomena in molecular clouds. 
\par
It has been theoretically predicted that cores develop in molecular clouds through gravitational contraction and/or external compression by turbulence and shocks (e.g., Mckee \& Ostriker 2007). 
If the (static) magnetic field plays a dominant role in supporting a core, the core will then evolve quasi-statically through the gravitationally induced drift of neutral particles with respect to ions and magnetic fields in a process known as ambipolar diffusion (Mestel \& Spitzer 1956). As neutral particles slowly migrate toward the center of the cloud, the mass-to-magnetic flux ratio gradually increases leading up to the onset of dynamical collapse. 
In the case of a weak and insignificant magnetic field, however, the core will readily initiate dynamical collapse if the internal non-magnetic support, i.e., thermal and turbulent pressure, is not sufficient. 
In order to investigate the stability and evolution of dense cores, it is therefore important to determine their physical properties, including information regarding the magnetic field. \par
If a uniform magnetic field is assumed at the initiation of contraction, the magnetic field lines frozen in the medium are expected to be dragged and distorted toward the center of the core by the contraction of the medium. The initially straight magnetic field lines are thus thought to distort into an hourglass-shaped structure under the influence of gravitation (e.g., Galli \& Shu 1993a,b).  
As contraction proceeds, even in the prestellar phase, a disk-like structure (a pseudo-disk) is expected to form in the center by mass accretion along the magnetic field lines, 
with the major axis of the disk aligned perpendicular to the direction of the magnetic field. These structures, if observed, are thus considered evidence of magnetically controlled star formation. \par
The direction of magnetic field lines projected onto the sky can be inferred from polarimetric observations of dust emission ($B$ $\perp$ $E$) and/or dichroic extinction ($B$ $\parallel$ $E$). According to the general alignment mechanism for interstellar dust grains, weakly magnetized and elongated dust grains spin about an axis (minor axis) oriented parallel to the magnetic field (e.g., Davis \& Greenstein 1951; Andersson, Lazarian, \& Vaillancourt 2015 for review). Dust emission polarimetry, particularly at sub-millimeter to far-infrared wavelengths, has proven to be a powerful technique for tracing magnetic field structures in regions of high (column) density, such as in giant molecular clouds or dark cloud complexes (e.g., Houde et al. 2004; Matthews et al. 2002), massive clumps/cores (e.g., Girart et al. 2009; Qiu et al. 2014), and protostellar envelopes (e.g., Girart et al. 2006; Rao et al. 2009). 
However, polarized dust emission from cold low-mass starless cores is too weak to be detected by presently available instruments. For this reason, magnetic field structures in starless cores have only been mapped toward dense central regions with a small number of data points (e.g., Ward-Thompson et al. 2000; Crutcher et al. 2004). There are thus few observations of starless core tracing and resolving magnetic fields from the center to the outermost regions. \par
In the present study, a magnetic field map of a molecular cloud core was constructed based on deep- and wide-field observations of dust extinction polarization at near-infrared (NIR) wavelengths. Extinction by dust is less severe at NIR wavelengths than at optical wavelengths (e.g., $\tau {}_{1.6 \mu {\rm m}} / \tau {}_{0.5 \mu {\rm m}}  \sim 0.18$), allowing the observations to penetrate the core and enabling the measurements of the dichroic polarization of light from a number of background stars. \par 
The molecular cloud core considered in the present study is FeSt 1-457, a nearby dense dark cloud core located close to the direction of the Galactic center (Feitzinger \& St\"{u}we 1984). The core is cataloged as a member of the cores of the Pipe Nebula (as core \#109 in Alves, Lombardi, \& Lada 2007; Onishi et al. 1999; Muench et al. 2007), a nearby dark cloud complex with a filamentary shape located at a distance of $d=130^{+24}_{-58}$ 
pc (Lombardi et al. 2006). Because of its relatively isolated geometry, simple shape, and the rich stellar field lying behind it, the density structure and other physical properties of FeSt 1-457 have been studied by measuring dust extinction (column density) of thousands of background stars at NIR wavelengths (Kandori et al. 2005). On the basis of the density structure determined using the Bonnor--Ebert sphere model (Bonnor 1956; Ebert 1955), FeSt 1-457 is estimated to have a radius of $R=18500$ AU (144$''$), a mass of $M_{\rm core}=3.55$ M$_{\odot}$, and a central density of $\rho_{\rm c}=3.5 \times 10^5$ cm$^{-3}$ at 130 pc. The turbulent motion in the core is small (approximately 30\%) compared with the thermal velocity dispersion at the gas kinetic temperature of 9.5 K in the core (Kandori et al. 2005; Rathborne et al. 2008). The results of a density structure study suggest that the core is no longer maintained stable against gravitational collapse by the thermal and small turbulent pressure (Kandori et al. 2005). No young stars possibly associated with the core have been found based on the NIR color of stars. 
Searches for young stellar objects (YSOs) using the Spitzer telescope (24 $\mu$m and 70 $\mu$m, Forbrich et al. 2009) have confirmed that the core is genuinely starless. These well-defined properties make FeSt 1-457 one of the most useful cores for studying the prestellar magnetic field structure using NIR polarimetry.
\section{Observations and Data Reduction}  
In the present study, FeSt 1-457 was observed using the $JHK_s$-simultaneous imaging camera SIRIUS (Nagayama et al. 2003) and its polarimetry mode SIRPOL (Kandori et al. 2006) on the IRSF 1.4-m telescope at the South African Astronomical Observatory (SAAO). IRSF/SIRPOL is one of the most powerful instruments for NIR polarization surveys, providing deep- (18.6 mag in the $H$ band, $5\sigma $ in one-hour exposure) and wide- ($7.\hspace{-3pt}'7 \times 7.\hspace{-3pt}'7$ with a scale of 0$.\hspace{-3pt}''$45 ${\rm pixel}^{-1}$) field polarization images. 
SIRPOL is a single-beam polarimeter. The uncertainty due to sky variation during exposures is typically $0.3\%$ in polarization degree. The uncertainty in polarization angle is less than 3$^{\circ}$ for bright sources (Kandori et al. 2006). 
Observations were made on the nights of 2006 June 15 and 16. The 10 sec exposures at four half-waveplate angles (in the sequence of $0^{\circ}$, $45^{\circ}$, $22.5^{\circ}$, and $67.5^{\circ}$) were performed at 10 dithered positions (one set). The total integration time was 1,900 seconds (19 sets) per waveplate angle. The typical seeing during the observations was approximately $1.\hspace{-3pt}''2$ (2.6 pixels) in the $H$ band. \par
The observed data were processed in the manner described by Kandori et al. (2006) using the Image Reduction and Analysis Facility of the National Optical Astronomy Observatory (NOAO/IRAF) and interactive data language (IDL) software (flat-field correction with twilight flat frames, median sky subtraction, and frame combine after registration). Software aperture polarimetry was carried out for a number of sources in the field of view. Point sources with a peak intensity greater than $10 \sigma $ above local sky background were detected. The local background was then subtracted from the original image using the mean of a circular annulus around the source on the image. This process was carried out for each position angle image ($I_{0}$, $I_{45}$, $I_{22.5}$, and $I_{67.5}$). The aperture radius was set to the full width at half maximum (FWHM) of the stars (2.6 pixels), and the sky annulus was set to 10 pixels with a width of 5 pixels. All sources with photometric errors greater than 0.1 mag were rejected. \par
The Stokes parameter for each star was obtained using $I = (I_{0} + I_{45} + I_{22.5} + I_{67.5})/2$, $Q = I_{0} - I_{45}$, and $U = I_{22.5} - I_{67.5}$. The polarization degree $P$ and polarization angle $\theta $ were then derived by $P = \sqrt{Q^2 + U^2}/I$ and $\theta = 0.5 {\rm atan}(U/Q)$.  
Debiasing of $P$ (Wardle \& Kronberg 1974) was not performed, because sources having a relatively large polarization degrees ($P_H \ge 1.5 \%$) and signal-to-noise ratio ($P_H / \delta P_H \ge 4$) were used for analysis (see Section 3.1). 
In the $H$ band data, polarizations for a total of 6,216 stars were measured. The present paper discusses the results in the $H$ band, where dust extinction is less severe than in the $J$ band, and the polarization efficiency is greater than in the $K_s$ band. 

\section{Results and Discussion} 
\subsection{Distortion of Magnetic Fields}
Figure 1 shows the $H$ band polarization vectors for stars superimposed on an intensity image taken for FeSt 1-457, which is located in the center of the image and appears as a dark nebula representing dust obscuration of background starlight. Although there are some kinks in the orientation of the polarization vector toward the core, the polarization degree ($P_{H}$) and angle ($\theta_{H}$) are relatively uniform throughout the region at a large distance from the core. \par 
The uniform ``off-core" component of the polarization ($P_{H,{\rm off}}$, $\theta_{H,{\rm off}}$) represents the effect of aligned dust particles along the line of sight but unrelated to the core. In this case, the values of $P_{H,{\rm off}}=3.9 \pm 0.8 \%$ and $\theta_{H,{\rm off}}=165^{\circ} \pm 10^{\circ}$ were estimated using 1,115 stars with $P_H /\delta P_H \ge 4$ located in the outside of the core radius (``off-core" region).
Note that the off-core polarization angle, $\theta_{H,{\rm off}}$, is in good agreement with the optical polarization data around FeSt 1-457 ($\theta_{\rm opt}=165^{\circ} \pm 4^{\circ}$, Franco et al. 2010). \par 
From Figures 2 and 3, the off-core component (solid line) is well aligned in $\theta_{H}$, whereas $P_{H}$ show a spread in its distribution. Spatial linear plane fitting of the off-core stars in $Q/I$ and $U/I$ was thus conducted. The distributions of $Q/I$ and $U/I$ values are independently modeled as $f(x,y)=A + Bx + Cy$, where $x$ and $y$ are the pixel coordinates and $A$, $B$, and $C$ are the fitting parameters. The distribution of the estimated off-core polarization vectors is shown in Figure 4. The off-core regression vectors are subtracted from the polarization vector measured for each star in order to determine the magnetic field associated with the core. After subtraction of the off-core polarization component, $P_{H}$ was reduced from 3.9$\%$ to 0.6$\%$, and $\theta_{H}$ became randomly distributed (Figures 2 and 3, dot-dashed line). The reduced polarization degree and the randomness of the polarization angle indicate the validity of the subtraction analysis. Note that the important conclusions do not change if the subtraction of uniform vector, i.e., (${P}_{H,{\rm off}}$, ${\theta}_{H,{\rm off}}$), is performed instead of the above plane fitting in $Q/I$ and $U/I$. \par 
The polarization distribution attributable solely to the effect of the core is shown in Figure 5. The polarization was successfully measured for 185 background stars located within the core radius (144$''$) and with a $P_{H}$ signal-to-noise ratio of greater than 4. 
Though the choice of the threshold in $P_H$ is somewhat arbitrary, a sufficient number of stars for further fitting analysis ($>100$) was obtained with a relatively high signal-to-noise ratio. 
The stars with small polarization degrees ($P_{H}$ $<$ 1.5$\%$) are not included in the results, because their polarization angles are sensitive to the subtraction analysis of the off-core polarization components. 
Note that the most of the off-core stars have residual polarization degrees of less than 1.5$\%$, as shown in Figure 2. 
\par
In Figure 5, the magnetic field follows a distinct axisymmetric shape reminiscent of an hourglass. The orientation of the magnetic field (magnetic axis) of the core is clearly in a north--south direction, providing angles that differed slightly from $\theta_{H,{\rm off}}$. The large number of polarization data points toward the core makes it possible to delineate the distorted magnetic field lines. The present results represent the first observational evidence of the hourglass-shaped magnetic fields in a starless core. 
The magnetic field geometry composed of the superposition of the distorted hourglass field and the uniform field is described in the Appendix. \par
The foreground stars do not affect the results, because the expected number of foreground stars toward the entire surface of FeSt 1-457 is less than one based on the Galaxy model (Wainscoat et al. 1992). The far-background stars also do not affect the results, because the radio observations of FeSt 1-457 (Kandori et al. 2005) showed that there is only a single gas component in this direction. 
\par
The existence of a distorted hourglass-shaped magnetic field should be interpreted as evidence for the mass condensation process. In observations, the curvature of the magnetic field lines for FeSt 1-457 is steep, particularly in the outer region. The curvature radius of the magnetic field lines is comparable to the core radius at the outermost region of the core. This means that the mass located around the outermost region of the core should move across a large distance in order to create the current distorted magnetic fields of the core. It is therefore obvious that the core radius was previously larger than the current radius and that the core contracted from far outside the current radius by dragging the frozen-in magnetic field lines. Since the core is in a nearly kinematically critical state, as shown in Section 3.3, the field distortion cannot be attributed to the dynamical collapse of the core. The observed distorted magnetic field structure is thus considered to be an imprint of the core formation process, in which mass was gathered and magnetic field lines were dragged to create the dense core. Note that the collapse of the core from the current stage could not create the steep curvature of the magnetic field lines in the outer regions of the core, because the collapse proceeds in a run-away fashion and the free-fall time is long in the diffuse outer region (see, e.g., Fiedler \& Mouschovias 1993). The steep magnetic curvature should thus be created mainly during the formation of the dense core. 
\par
It is noteworthy that the present observational data successfully trace the dust polarization into the dense part of the core because of the linearity of the relationship between the dust extinction (i.e., reddening of the stellar color) and the polarization degree (Figure 6). It was found that there is no turnover in the relationship up to $A_V \approx 15$ mag, with a slope of $P_{H} / (H-K_{s}) \approx 4.8$ \% mag$^{-1}$. This demonstrates that the polarization of background starlight can be used to trace the magnetic field inside cold and dense clouds, despite reports to the contrary for some dark clouds (e.g., Goodman et al. 1995; Arce et al. 1998). \par
In contrast to the present result, Alves et al. (2014a,b) reported the existence of a kink at $A_V \approx 10$ mag on the $A_V$ vs. $P_{H} / A_V$ diagram for FeSt 1-457. They also reported that $P_{H} / A_V$ is a decreasing function of $A_V$. Their results, however, consider the superposition of polarization both from the core and the off-core medium, which have different position angles. Note that the sum of vectors with different position angles can produce a depolarization effect. We obtained a $A_V$ vs. $P_{H} / A_V$ diagram similar to that obtained by Alves et al. (2014a,b) by using the present polarization data before the subtraction of off-core components. On the basis of the \lq \lq pre-subtraction'' data, $P_{H} / A_V$ clearly decreases with increasing $A_V$, although the slope of the correlation is slightly different from that obtained by Alves et al. (2014a,b). A detailed study of the relationship between polarization and extinction in FeSt 1-457 will be presented in a subsequent paper. \par
Alves et al. (2014a,b) and Ju\'{a}rez et al. (2016) reported sub-millimeter (submm) dust continuum polarimetry at 345 GHz for FeSt 1-457. A ``polarization hole'' can clearly be observed in their submm polarization data. The maximum and minimum polarization degree were measured for the diffuse outermost region of the core and the densest center, respectively, (anti-correlation between $P_{\rm submm}$ and $A_V$). It is therefore obvious that the magnetic field geometry of the core inferred from the submm polarizations does not trace the overall magnetic properties of the core. In the present study, we therefore focus on the present NIR polarization data in order to discuss the magnetic field for FeSt 1-457. 
\subsection{Parabolic Modeling}
The most probable configuration of magnetic field lines pervading the core, estimated using a parabolic function and its rotation, is shown in Figure 5 (solid white lines). 
The coordinate origin for the parabolic function is fixed to the center of the core measured on the extinction map (R.A.~=~17$^{\rm h}$35$^{\rm m}$47$.\hspace{-3pt}^{\rm s}$5, Decl.~=~$-25^{\circ}$32$'$59$.\hspace{-3pt}''0$, J2000; Kandori et al. 2005). The fitting parameters are $\theta_{\rm mag}=179^{\circ} \pm 11^{\circ}$ and $C = 1.04 (\pm 0.45) \times 10^{-5}$ ${\rm pixel}^{-2}$ ($= 5.14 \times 10^{-5}$ ${\rm arcsec}^{-2}$)  
for the parabolic function $y = g + gC{x^2}$, where $g$ specifies the magnetic field lines, $\theta_{\rm mag}$ is the position angle of the magnetic field direction or the $x$-axis, which is measured from the north through east, and $C$ determines the degree of curvature of the parabolic function. 
The observational error for each star was taken into account in the calculations of $\chi^2 (= \sum_{i=1}^n (\theta_{\rm obs,{\it i}} - \theta_{\rm model}(x_i,y_i))^2 / \delta \theta_i^2$, where $n$ is the number of stars; $x$ and $y$ are the coordinates of the $i$th star; $\theta_{\rm obs}$ and $\theta_{\rm model}$ are the polarization angles from the observations and the model, respectively; and $\delta \theta_i$ is the observational error) in the fitting procedure. \par
The parabolic fitting appears to be reasonable because the standard deviation of the residual angles $\theta_{\rm res} = \theta_{\rm obs} - \theta_{\rm fit}$, where $\theta_{\rm fit}$ is the best fit $\theta_{\rm model}$, is smaller for the parabolic function ($\delta \theta_{\rm res} = 10.24^{\circ}$, Figure 7) than for the uniform-field case ($16.25^{\circ}$). 
In order to confirm this statistically, the uncertainty in $\delta \theta_{\rm res}$ was estimated by using the bootstrap method. A random number following the normal distribution with the same width as the observational error was added in each star, and fitting with the parabolic function was conducted. This process was repeated 1,000 times in order to obtain the dispersion of the resulting $\delta \theta_{\rm res}$ distribution. Values of $0.89^{\circ}$ and $0.70^{\circ}$ were obtained for the uncertainties in $\delta \theta_{\rm res}$ and the uniform-field case, respectively. It is now clear that the value of $\delta \theta_{\rm res} (10.24^{\circ} \pm 0.84^{\circ})$ is statistically distinguishable from that in the uniform-field case ($16.25^{\circ} \pm 0.70^{\circ}$). Therefore, the existence of the distorted field is most likely real. \par
In Figure 5, the distorted magnetic field shape is much more apparent in the right-hand side of the core than in the left-hand side. The parabolic fitting results might be affected by the polarization vectors in the right-hand side of the core. In order to confirm this, the 100 polarization vectors located in the right half of the core were masked, and parabolic fitting was conducted by using the remainder of the vectors. Best-fit values of $\theta_{\rm mag}=177^{\circ}$ and $C=5.00 \times 10^{-6}$ pixel$^{-2}$ were obtained. These values are consistent with those obtained without a mask, and it is clear that the field distortion exists in both sides of the core. \par
The intrinsic dispersion $\delta \theta_{\rm int} = (\delta \theta_{\rm res}^2 - \delta \theta_{\rm err}^2)^{1/2}$ estimated using the parabolic fitting was found to be $6.90^{\circ}\pm2.72^{\circ}$ ($0.12\pm0.048$ radian), where $\delta \theta_{\rm err}$ is the standard deviation of the observational error in the polarization measurements. Note that $\delta \theta_{\rm int}$ serves as the upper limit of the intrinsic dispersion because a better choice of the function than that used in the present case (i.e., parabolic form) can reduce the dispersion in the residual angles. \par
Assuming that the magnetic field is frozen in the medium, the intrinsic dispersion $\delta \theta_{\rm int}$ of the magnetic field direction can be attributed to the Alfven wave perturbed by turbulence. The strength of the plane-of-sky magnetic field (${B}_{\rm pos}$) can be estimated from the relation ${B}_{\rm pos} = {C}_{\rm corr} (4 \pi \rho)^{1/2} \sigma_{\rm turb} / \delta \theta_{\rm int}$, where $\rho$ and $\sigma_{\rm turb}$ are the mean density of the core and the turbulent velocity dispersion, respectively (Chandrasekhar \& Fermi 1953), and ${C}_{\rm corr}$ is a correction factor whose inclusion has been suggested by theoretical studies. In the original formulation, ${C}_{\rm corr} = 1$ (Chandrasekhar \& Fermi 1953), whereas in the present study, a value of ${C}_{\rm corr} = 0.5$ (Ostriker et al. 2001) was adopted. 
Using the data of the mean density ($7.93 \pm 2.51 \times 10^{-20}$ g cm$^{-3}$) and the turbulent velocity dispersion ($0.0573 \pm 0.006$ km s${}^{-1}$) from Kandori et al. (2005) and the value of $\delta \theta_{\rm int}$ derived in the present study, a relatively weak magnetic field was obtained as the lower limit for the total field strength ($|B|$): $B_{\rm pos} = 23.8 \pm 12.1$ (${C}_{\rm corr}/0.5$)($d/130$ ${\rm pc}$)$^{-1/2}$ $\mu {\rm G}$. Note that this value is the mean plane-of-sky field strength for the core. \par
In the error estimation, the uncertainties in the mean density, turbulent velocity dispersion, and $\delta \theta_{\rm int}$ are considered. The turbulent velocity dispersion value used here is based on the line width in N$_2$H$^+$ ($J=1-0$) toward the center of the core (Kandori et al. 2005). The radial distribution of the line width is confirmed to be flat within the radius of $70''$ (see also Aguti et al. 2007), and the mean turbulent velocity dispersion for the core was therefore set to the same value as that toward the core center. The difference in the velocity dispersion between the central spectrum and the composite spectrum of the nine central positions was used as the uncertainty in the turbulent velocity dispersion. 
\subsection{Stability and Evolutionary Status of the Core}
Given the detailed magnetic field structure and other physical properties now known for FeSt 1-457, this starless core can serve as a good laboratory for studying the initial conditions of star formation. The ability of the magnetic field to support the core against gravity was investigated using the parameter ${\lambda} = ({M}/{\Phi})_{\rm obs} / ({M}/{\Phi})_{\rm critical}$, which represents the ratio of the observed mass-to-magnetic flux to a critical value, $1/2\pi {G}^{1/2}$, suggested by theory (Mestel \& Spitzer 1956; Nakano \& Nakamura 1978). A value of ${\lambda} \approx 2.0$ (magnetically supercritical) was derived in the present case. 
The critical mass-to-flux ratio for the case of a disk geometry was used in the present study. The numerical coefficient is slightly different for the spherical geometry case ($2/3\pi {G}^{1/2}$: Strittmatter 1966). \par
Although the theoretical magnetic critical mass $M_{\rm mag} = 1.77$ ${\rm M}_{\rm \odot}$ for the core is sufficiently small in comparison with the observed mass ($M_{\rm core} = 3.55$ ${\rm M}_{\rm \odot}$), this does not necessarily mean that the core is in a dynamical collapse state. In order to evaluate the stability, the critical mass for the core can be considered as $M_{\rm cr} \simeq M_{\rm mag}+M_{\rm BE}$ (Mouschovias \& Spitzer 1976; Tomisaka, Ikeuchi, \& Nakamura 1988; McKee 1989), where $M_{\rm BE}$ is the Bonnor--Ebert mass (Bonnor 1956; Ebert 1955). Thermal support is also important when considering the stability of low-mass clouds, and the approximation of $M_{\rm cr}$ is accurate to within approximately 5\% for $M_{\rm core} \simlt 8 M_{\rm mag}$ (McKee 1989). The Bonnor--Ebert mass calculated from the effective sound speed and the external pressure of the core is $M_{\rm BE} = 1.19$ ${\rm M}_{\rm \odot}$ (Kandori et al. 2005). The contribution from both the thermal and turbulent pressure is included in the mass via the effective sound speed. The obtained critical mass of the core, $M_{\rm cr}=2.96$ ${\rm M}_{\rm \odot}$ is comparable to $M_{\rm core}$, suggesting a nearly critical state. \par
FeSt 1-457 appears to be not far from the critical state just before or after the onset of dynamical collapse to form star(s). Radio molecular line observations have not detected a signature of gas infalling motion in the core, and furthermore, possible oscillating gas motion in the outer shell of the core has been reported (Aguti et al. 2007). The kinematic status of the core is consistent with the present results for the nearly critical feature of the core. It is noteworthy that the magnetic or thermal/turbulent pressure alone cannot support the core. Further diffusion of magnetic fields and/or turbulent dissipation would lead to the onset of the dynamical collapse of the core. 
Note that this is based on a picture of the relatively isolated star formation. External compression of the core (e.g., Frau et al. 2015 for the possibility of the cloud-cloud collision in the Pipe Nebula) may also initiate the star formation in the core. \par 
Rotation of the core does not affect its stability. The rotational kinetic energy of the core is only approximately $1$\% of the gravitational energy (Aguti et al. 2007). Uncertainty arising from the use of the plane-of-sky field component increases the importance of magnetic support. The magnetic field close to a pole-on geometry leads to the condition $M_{\rm cr} > M_{\rm core}$ (subcritical). Therefore, 3D modeling of the polarization structure is desirable in order to determine the line-of-sight inclination angle of the core's magnetic axis. \par
%
The formation of a nearly critical core accompanied by a distorted magnetic field is an open problem. Consideration of the interplay among gravity, thermal pressure, and turbulence is necessary in order to understand the formation and evolution mechanism of FeSt 1-457. Quasi-static contraction under the support of a magnetic field (Shu 1977) requires a strong magnetic field, and this is not the case for FeSt 1-457 ($\lambda \approx 2$, magnetically supercritical). In the other extreme, supersonic turbulence can produce cores that collapse dynamically and that are accompanied by highly supersonic infalling motion (e.g., Mac Low \& Klessen 2004). This is also not the case for FeSt 1-457 because nearly critical stability was shown in the present study and radio observations (Kandori et al. 2005; Aguti et al. 2007) showed quiescent kinematic gas motion in the core. Core formation between the two extreme models may account for the formation of FeSt 1-457 (e.g., Nakamura \& Li 2005; Basu et al. 2009a,b). Further theoretical studies on such a case are desirable. \par
The relative importance of the magnetic field for the support of the core is investigated through the ratios of the thermal and turbulent energy to the magnetic energy, which are expressed as ${\beta} \equiv 3{C}_{\rm s}^{2}/{V}_{\rm A}^{2}$ and ${\beta}_{\rm turb} \equiv {\sigma}_{\rm turb,3D}^{2}/{V}_{\rm A}^{2}=3 {\sigma}_{\rm turb,1D}^{2}/{V}_{\rm A}^{2}$, where ${C}_{\rm s}$, ${\sigma}_{\rm turb}$, and ${V}_{\rm A}$ denote the isothermal sound speed at 9.5 K, the turbulent velocity dispersion, and the Alfven velocity, respectively. The numerical coefficients differ in the equation depending on whether the velocity dispersion is in 1D or 3D. These ratios were found to be ${\beta} \approx 1.88$ and ${\beta}_{\rm turb} \approx 0.17$. 
%
%
FeSt 1-457 is dominated by a static magnetic field and thermal support with a small contribution from turbulence. Turbulence could be an important parameter in the formation and development of the core, although it seems to have already mostly dissipated at the observed stage in the prestellar evolution. Therefore, a static magnetic field, the predominant component of anisotropic pressure in the core, should act as a key parameter controlling the star formation process in FeSt 1-457. \par
Figure 8 shows the relationships among the orientations (position angles) of the axis of the elongation of the core (${\rm \theta}_{\rm elon} \approx 90^{\circ}$), the rotation axis (${\rm \theta}_{\rm rot} = 140^{\circ}$--$160^{\circ}$, Aguti et al. 2007), and the magnetic field axis (${\rm \theta}_{\rm mag} = 179^{\circ}$). Note that the elongation of the core in the outermost region is roughly perpendicular to the rotation axis. As shown in the figure, the orientation of the elongated (disk-like) structure in the central region is clearly perpendicular to the magnetic axis. This geometrical relationship supports the occurrence of mass accretion along magnetic field lines, as suggested by previously reported theories (e.g., Galli \& Shu 1993a,b) or the magneto-hydrostatic configuration (e.g., Tomisaka, Ikeuchi, \& Nakamura 1988). For $M_{\rm mag} > M_{\rm BE}$, a magneto-hydrostatic configuration can produce a flattened inner density distribution with its major axis perpendicular to the direction of the magnetic field lines (Tomisaka, Ikeuchi, \& Nakamura 1988), which is consistent with the observed magnetic geometry and extinction distribution of the core. There is a slight misalignment between the rotational and magnetic axes. The initial angular momentum of the core was likely not well aligned with the magnetic axis. Note that the orientation of the magnetic axis is consistent with that of the large parsec-scale magnetic field (Alves \& Franco, 2008), and both are roughly perpendicular to the filament axis of the Pipe Nebula, which is indicative of the formation of the magnetically controlled structure on both large and small scales. 
\\
\par
The present study reveals the stability and evolutionary status of the starless core FeSt 1-457. Both the magnetic and thermal properties of the core suggest that the stability of the core is in a nearly critical state, although the core is magnetically supercritical. The hourglass-shaped magnetic field around FeSt 1-457 is indicative of the core formation history, as suggested by its steep magnetic curvature. 
Since magnetic or thermal/turbulent pressure alone cannot support the core, it was concluded that further diffusion of the magnetic field and/or turbulent dissipation would lead to the onset of dynamical collapse. 
On the basis of these observational results, FeSt 1-457 can be regarded to be in the earliest stage of collapse, i.e., a stage before the formation of protostar(s) and the development of the (supersonic) infalling gas motion, following the attainment of the supercritical condition. These properties would make FeSt 1-457 one of the best prestellar cores for future studies of the initial condition of star formation.  
\subsection*{Acknowledgements}
We are grateful to Shuji Sato, Tomoaki Matsumoto, Masao Saito, Shogo Nishiyama, and Mikio Kurita for their helpful comments and suggestions. Thanks are due to the staff at SAAO for their kind help during the observations. We would also like to thank Tetsuo Nishino, Chie Nagashima, and Noboru Ebizuka for their support in the development of SIRPOL and its calibration and its stable operation with the IRSF telescope. The IRSF/SIRPOL project was initiated and supported by Nagoya University, National Astronomical Observatory of Japan, and the University of Tokyo in collaboration with the South African Astronomical Observatory under a financial support of Grants-in-Aid for Scientific Research on Priority Area (A) No. 10147207 and No. 10147214, and Grants-in-Aid No. 13573001 and No. 16340061 of the Ministry of Education, Culture, Sports, Science, and Technology of Japan. RK, MT, NK, and KT (Kohji Tomisaka) also acknowledge additional support through Grants-in-Aid Nos. 16077101, 16077204, 16340061, 21740147, 26800111, 16K13791, and 15K05032. 
\subsection*{Appendix: Magnetic Field Geometry}
In the main text, we used the terms ``hourglass-shaped" and ``parabolic" as the same meaning to represent the shape of the magnetic field. Magnetic fields solely associated with dense cores, as demonstrated by theoretical simulations, can be approximated by a parabolic function, with terms for both the field lines and the curvature to express the distortion characteristics. 
Hourglass-shaped magnetic fields, which have a uniform field component that is much stronger than that in the parabolic case, have been observed in star-forming regions (e.g., Tamura et al. 1987). Furthermore, integral- or S-shaped magnetic fields have also been observed under similar conditions (e.g., Figure 6 of Vall\'{e}e 2003). \par
These three types of magnetic geometry can be explained by the superposition of a parabolic field and a parallel uniform field or a uniform field inclined on the plane of the sky. An example of a parabolic case is shown in Figure S1(a). The curvature in this case is $C=1.0 \times 10^{-4}$ ${\rm pixel}^{-2}$ 
and the polarization degree is proportional to the amount of dust calculated based on the structure of the Bonnor--Ebert sphere. The circle indicates the boundary of the core. Figure S1(b) shows the parabolic field superimposed on the parallel uniform field, yielding an hourglass-shaped field, and Figure S1(c) shows the parabolic field superimposed on an inclined uniform field, yielding an integral-shaped field. \par
In the present study, integral-shaped magnetic fields were observed for FeSt 1-457, as shown in Figure 1. The ambient (uniform) off-core field was slightly inclined ($\theta_{\rm off}=165^{\circ}$) with respect to the magnetic axis of the core ($\theta_{\rm mag}=179^{\circ}$). After subtraction of the off-core polarization components, the parabolic field shown in Figure 5 was obtained.

\clearpage 

\begin{figure}[t]  
\begin{center}
 \includegraphics[width=6.5 in]{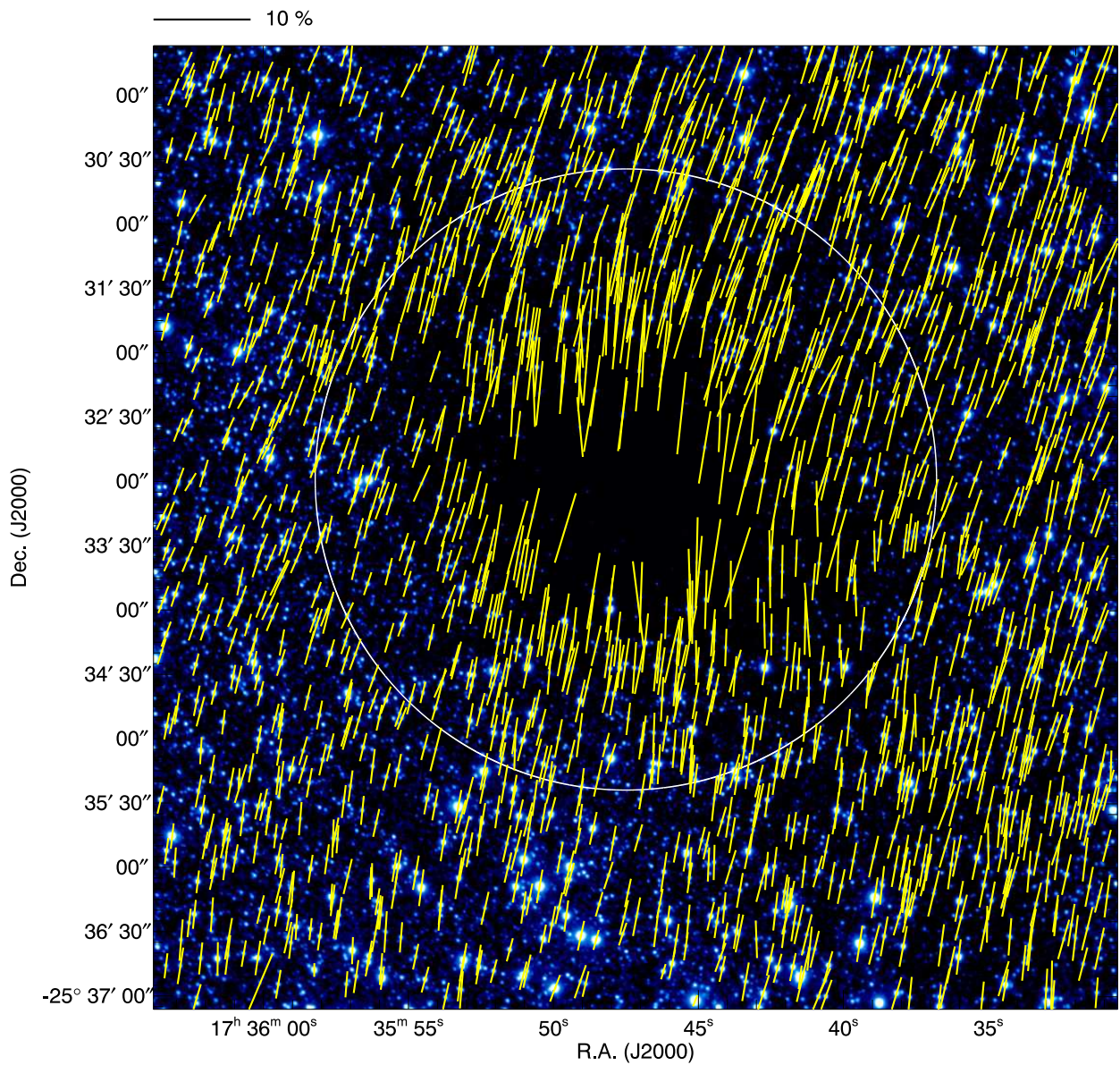}
 \caption{Polarization vectors of point sources superimposed on an intensity image in the $H$ band. The core radius (144$''$) is indicated by the white circle. The scale of the 10$\%$ polarization degree is shown at the top.}
   \label{fig1}
\end{center}
\end{figure}

\clearpage 

\begin{figure}[t]  
\begin{center}
 \includegraphics[width=4.5 in]{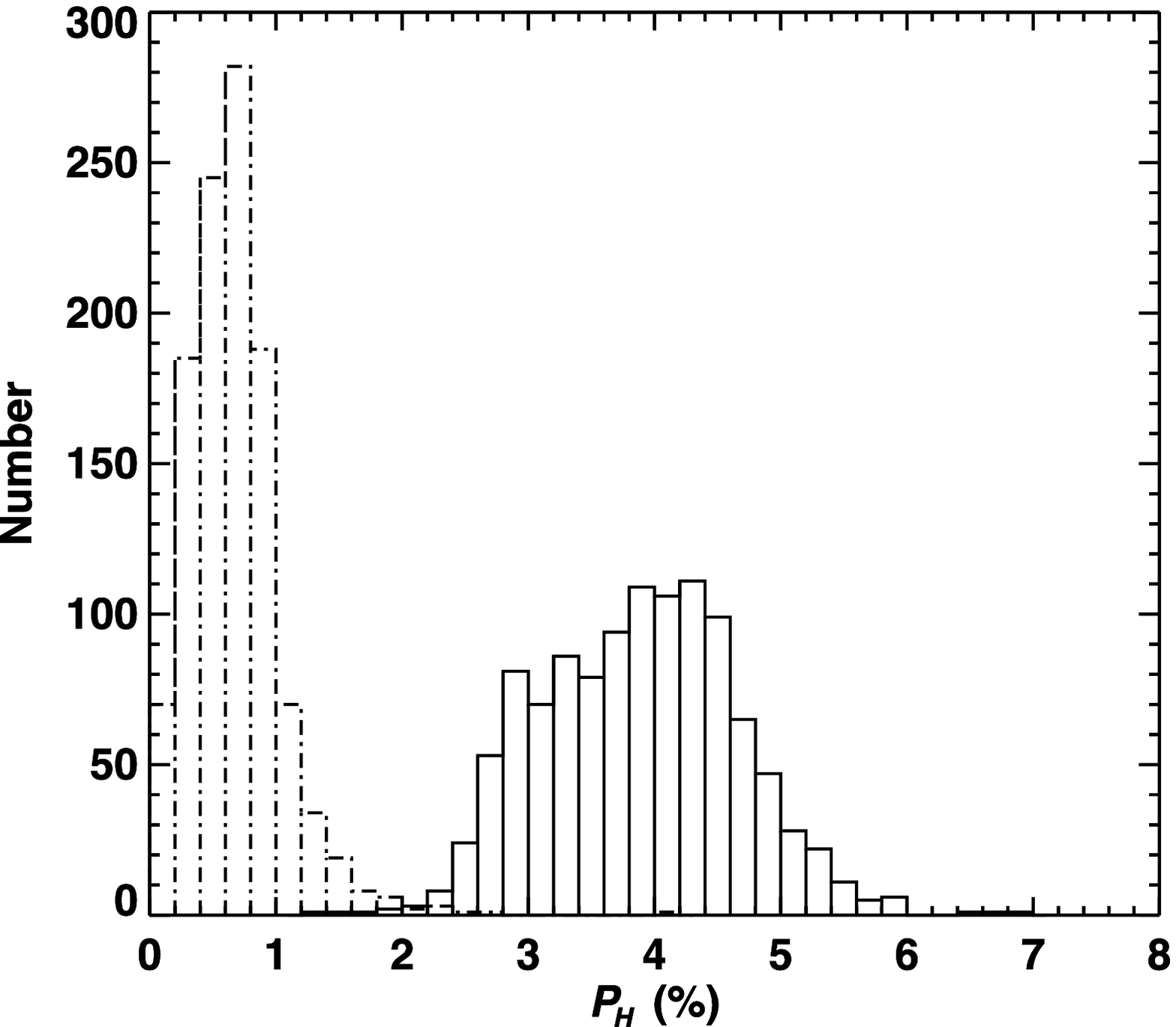}
 \caption{Histogram of $P_{H}$ for stars in the off-core region before (solid line) and after (dot-dashed line) subtraction of the off-core component.}
   \label{fig2}
\end{center}
\end{figure}

\begin{figure}[t]  
\begin{center}
 \includegraphics[width=4.5 in]{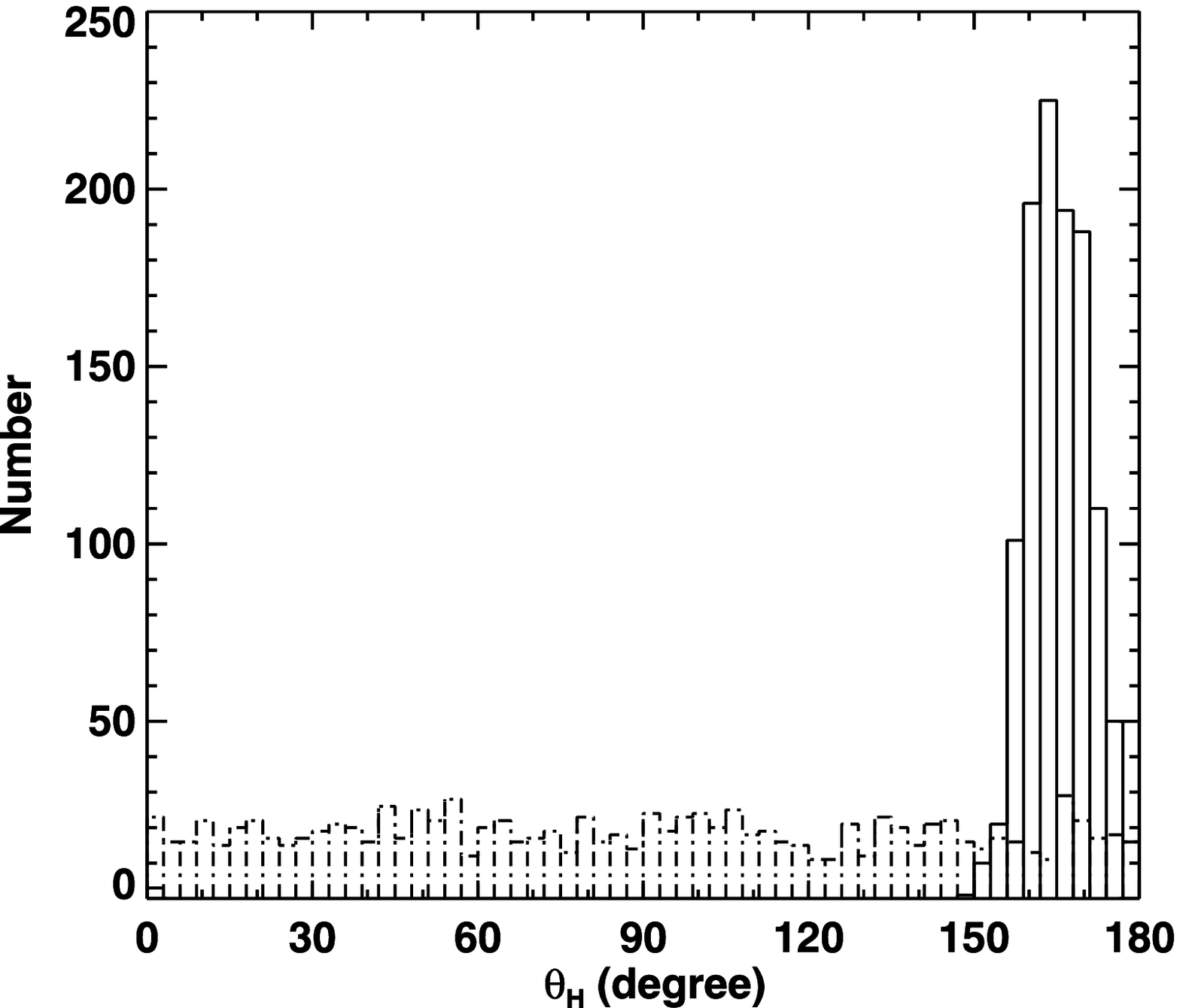}
 \caption{Histogram of $\theta_H$ for stars in the off-core region before (solid line) and after (dot-dashed line) subtraction of the off-core component.}
   \label{fig2}
\end{center}
\end{figure}

\clearpage 

\begin{figure}[t]  
\begin{center}
 \includegraphics[width=6.5 in]{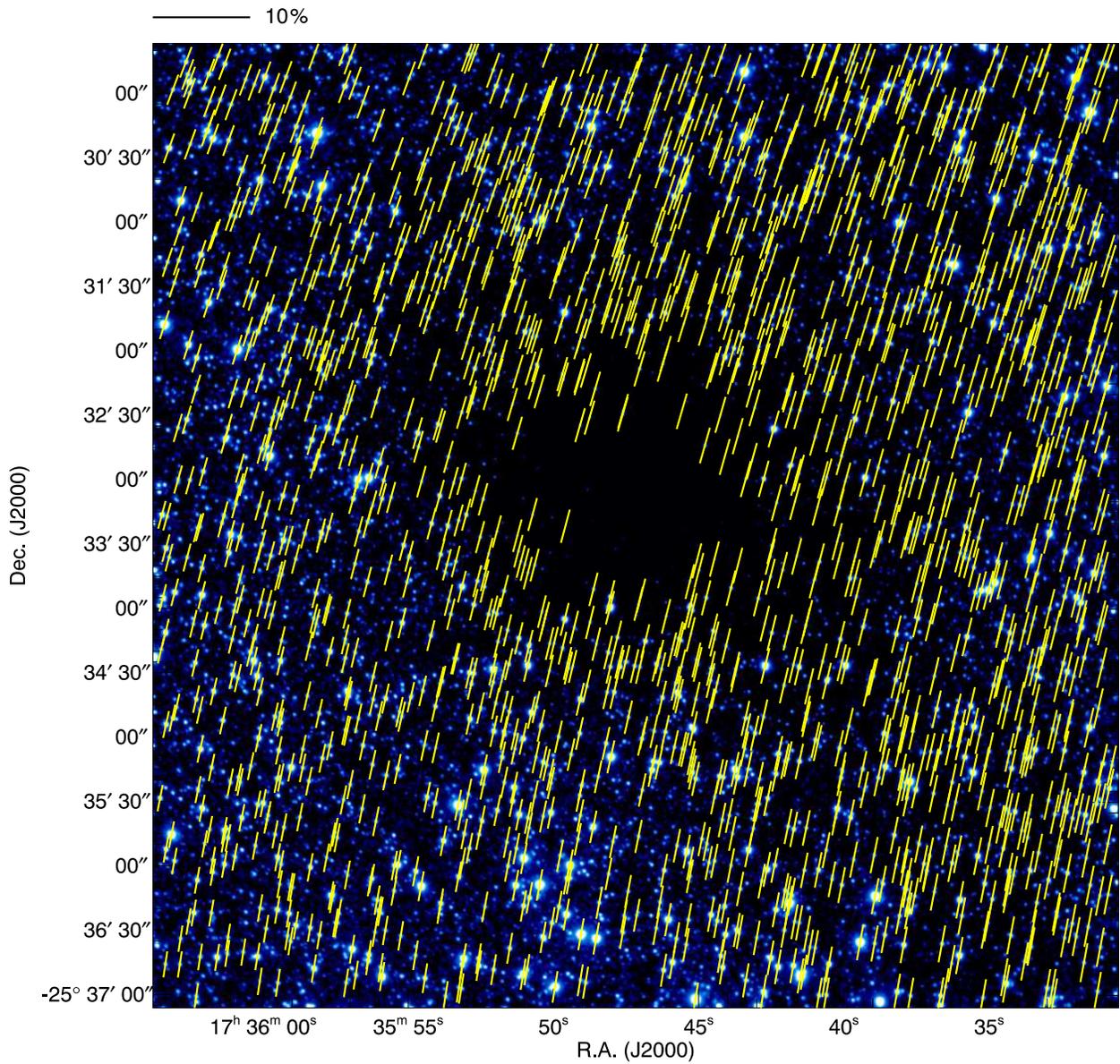}
 \caption{Estimated off-core polarization vectors superimposed on an intensity image in the $H$ band. Note that these vectors are not obtained directly from observations. Off-core vectors estimated by fitting are plotted at the position of each star. The background image is the same as that for Figure 1. The scale of the 10$\%$ polarization degree is shown at the top.}
   \label{fig1}
\end{center}
\end{figure}

\clearpage 

\begin{figure}[t]  
\begin{center}
 \includegraphics[width=6.5 in]{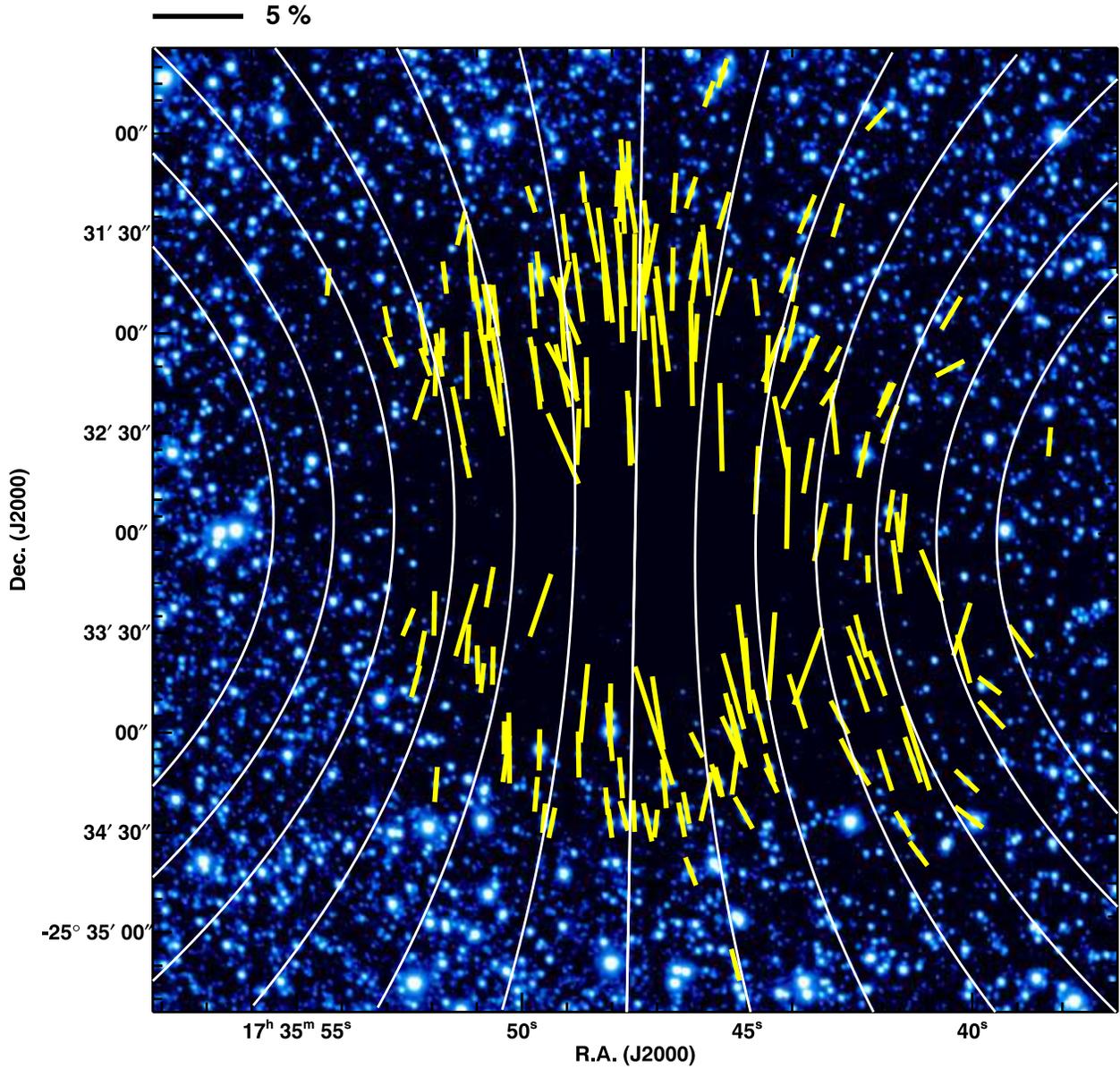}
 \caption{Polarization vectors after subtraction of the off-core component. The field-of-view is 288$''$ or 0.19 pc at a distance of 130 pc, which is equal to the diameter of the core. The background image for this figure is the a zoomed-in version of that used in Figures 1 and 4. The white lines indicate the direction of the magnetic field inferred from parabolic fitting. The scale of the 5$\%$ polarization degree is shown at the top.
}
   \label{fig3}
\end{center}
\end{figure}

\clearpage 

\begin{figure}[t]  
\begin{center}
 \includegraphics[width=6.5 in]{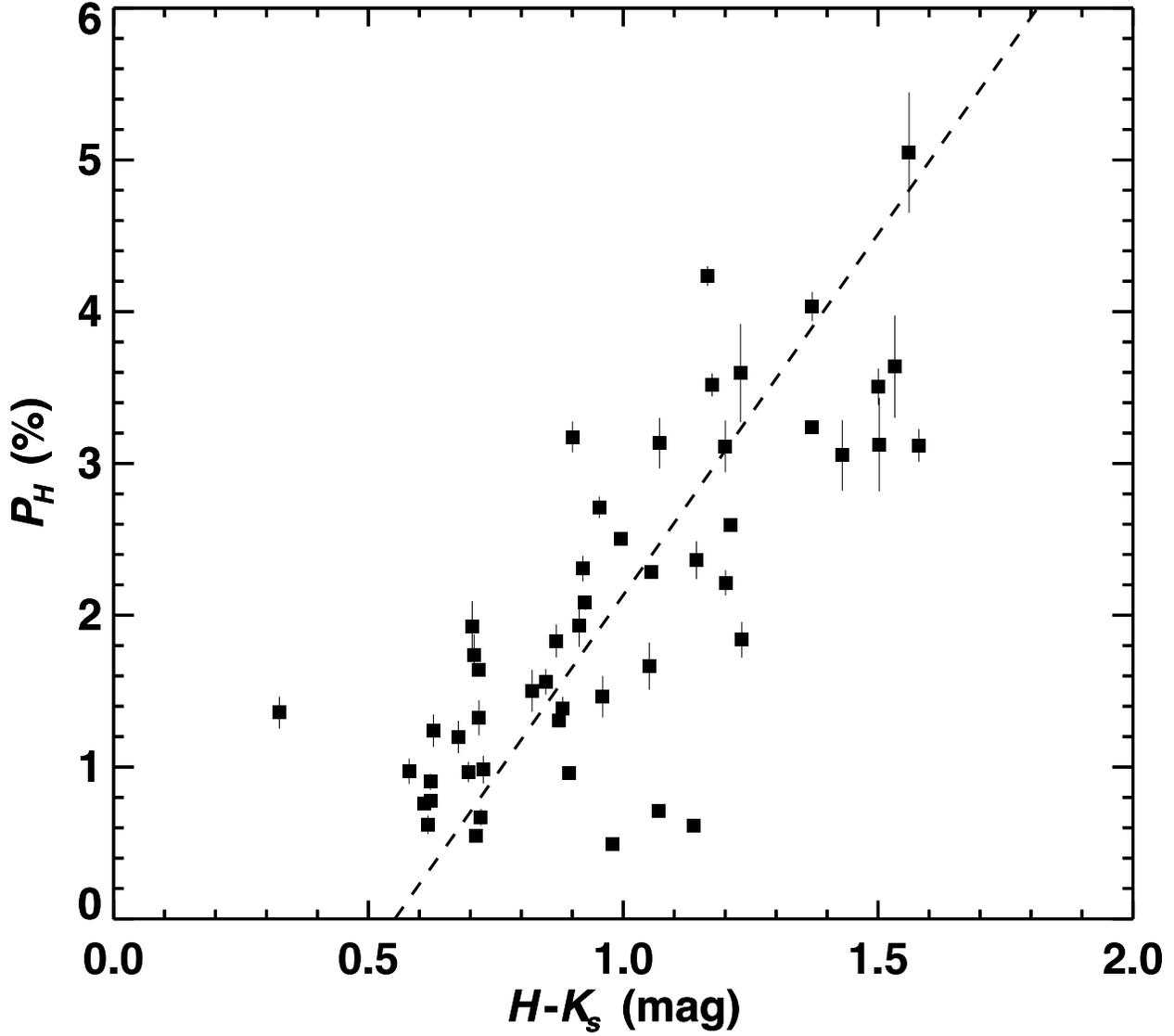}
 \caption{Relationship between $H-K_{s}$ color and polarization degree toward background stars with $P_{H}$ taken after the subtraction of the off-core component. Only stars with $P_H / \delta P_H \geq 10$ are plotted.}
   \label{fig4}
\end{center}
\end{figure}

\clearpage 

\begin{figure}[t]  
\begin{center}
 \includegraphics[width=6.5 in]{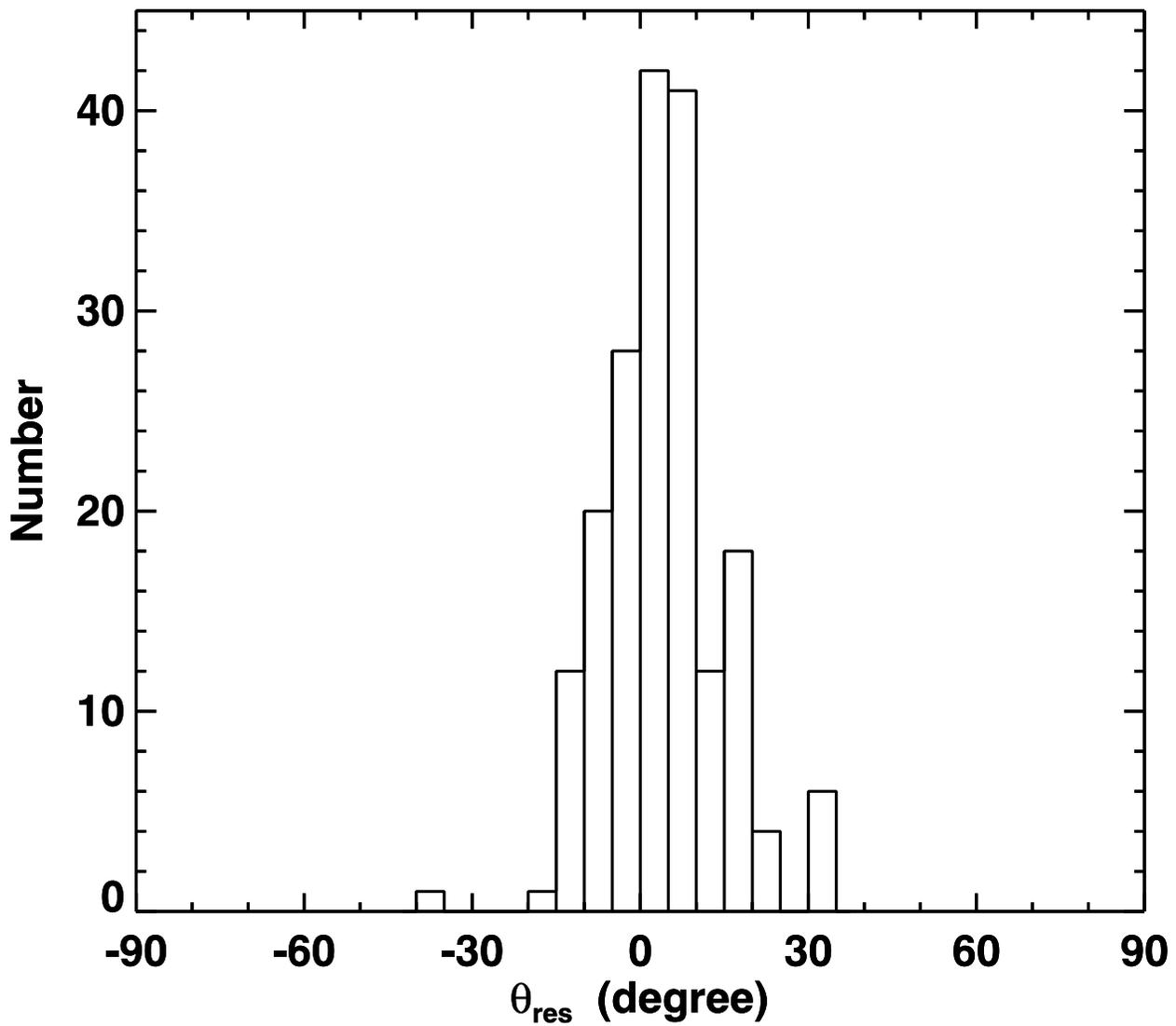}
 \caption{Histogram of the residual of the observed polarization angle $\theta_{\rm res}$ after subtraction of the angle obtained by parabolic fitting.}
   \label{fig6}
\end{center}
\end{figure}

\clearpage 

\begin{figure}[t]  
\begin{center}
 \includegraphics[width=6.5 in]{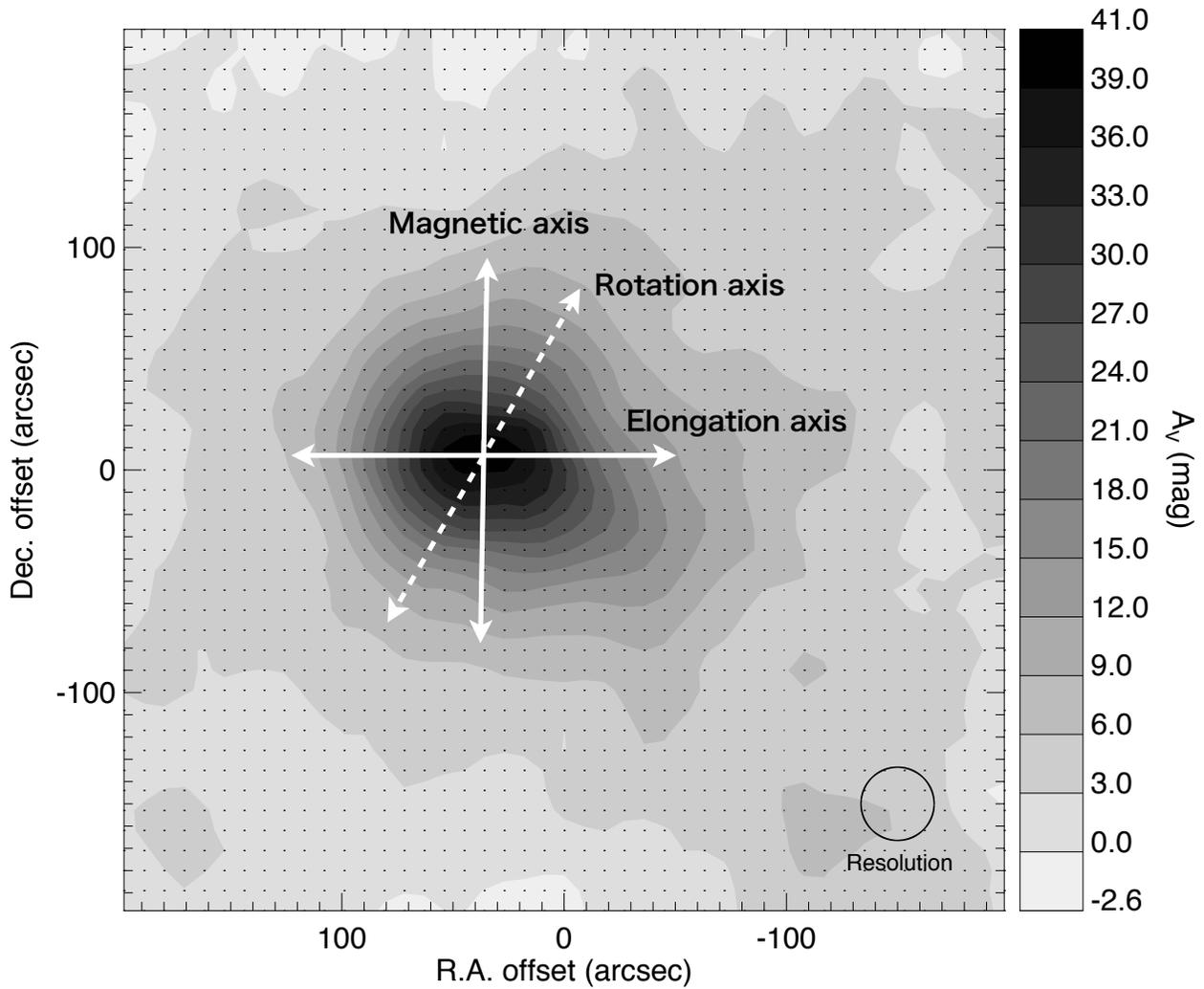}
 \caption{Relationships among the axis of elongation of the core ($\theta_{\rm core} \approx 90^{\circ}$), the rotation axis ($\theta_{\rm rot} = 140^{\circ}$--$160^{\circ}$), and the magnetic field axis ($\theta_{\rm mag} = 179^{\circ}$) superimposed on the dust extinction ($A_V$) map taken from Kandori et al. (2005). Note that the elongation of the core in the outermost region is roughly perpendicular to the rotation axis. The step of the $A_V$ contour is 3 mag. A value of $A_V = 41$ mag was found near the center. The center of the map is (R.A., Decl.)$=$(17$^{\rm h}$35$^{\rm m}$45$.\hspace{-3pt}^{\rm s}$0, $-25^{\circ}$33$'$11$.\hspace{-3pt}''0$, J2000), and the resolution of the $A_V$ map is $30''$. The small dots on the map are the measurement points of extinction.}
   \label{fig5}
\end{center}
\end{figure}

\clearpage 

\beginsupplement

\clearpage 

\begin{figure}[t]  
\begin{center}
 \includegraphics[width=6.5 in]{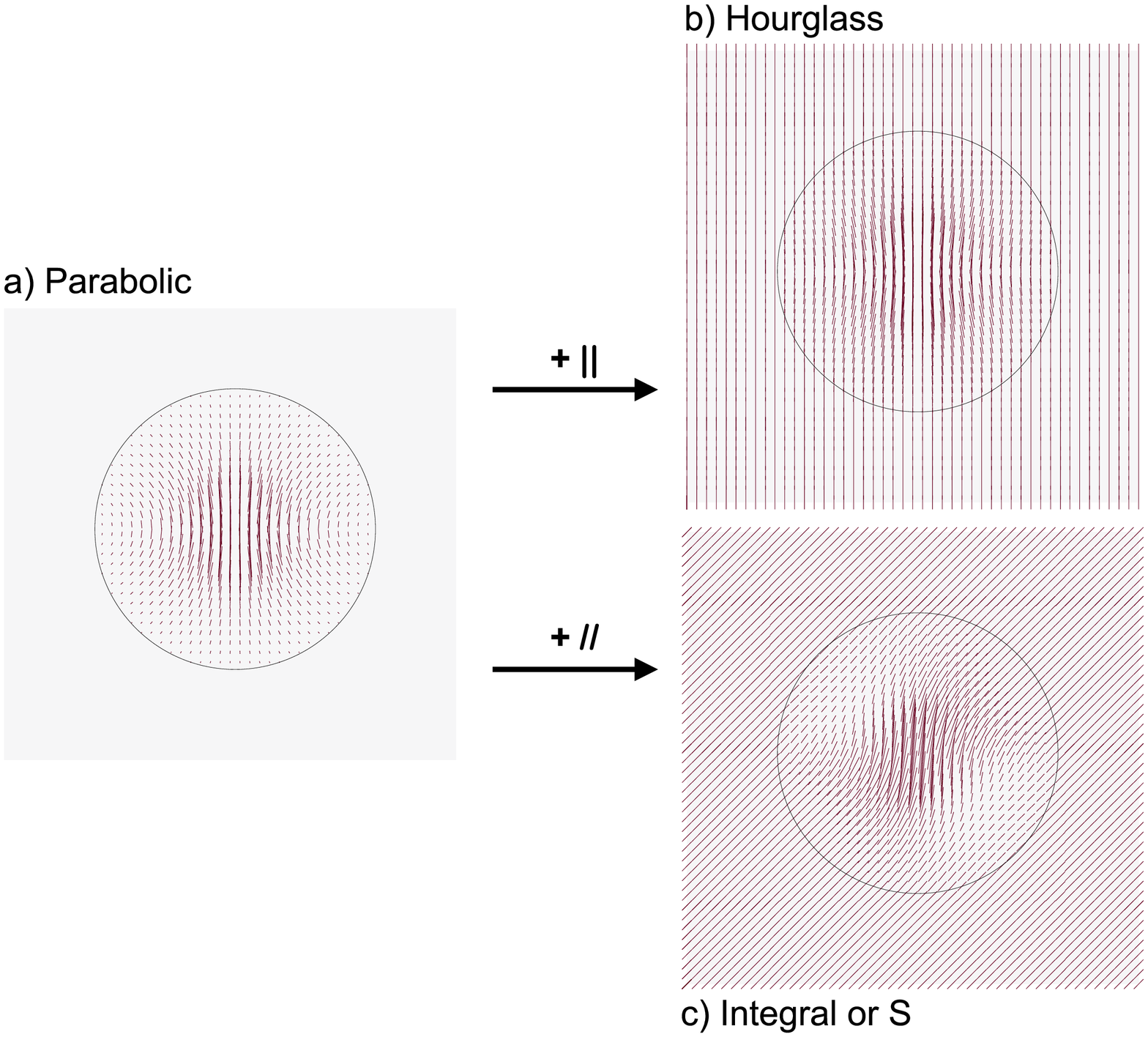}
 \caption{Schematic of the three dense core magnetic field geometries.}
   \label{fig5}
\end{center}
\end{figure}

\end{document}